\documentclass{JAC2003}

\usepackage{graphicx}
\usepackage{dcolumn}
\usepackage{bm}
\usepackage{url}
\usepackage{amsmath,amssymb}
\usepackage{verbatim}  
\usepackage{relsize}
\usepackage{textcomp}
\usepackage{booktabs}
\usepackage{multirow}
\usepackage{wasysym} 
\usepackage{xspace}  
\usepackage{fixltx2e}  
\usepackage{multirow}  

\newcommand{\bs}{$\beta^*$\xspace}

\newcommand{\sig}{$\sigma$\xspace}

\newcommand{\um}{$\mathrm{\mu}$m\xspace}
\newcommand{\urad}{$\mathrm{\mu}$rad\xspace}

\begin{document}
\title{BASELINE LHC MACHINE PARAMETERS AND CONFIGURATION \\ OF THE 2015 PROTON RUN}

\author{R. Bruce\thanks{roderik.bruce@cern.ch}, G. Arduini, S. Fartoukh, M. Giovannozzi, \\M. Lamont, E. Metral, T. Pieloni, S. Redaelli, J. Wenninger \\CERN, Geneva, Switzerland}

\maketitle

\begin{abstract}

 This paper shows the baseline LHC machine parameters for the 2015 start-up. Many systems have been upgraded during LS1 and in 2015 the LHC will operate at a higher energy than before and with a tighter filling scheme. Therefore, the 2015 commissioning phase risks to be less smooth than in 2012. The proposed starting configuration puts the focus on feasibility rather than peak performance and includes margins for operational uncertainties. Instead, once beam experience and a better machine knowledge has been obtained, a push in \bs and performance can be envisaged. In this paper, the focus is on collimation settings and reach in \bs---other parameters are covered in greater depth by other papers in these proceedings. 
\end{abstract}

\section{INTRODUCTION}
The first running period of the LHC, Run~I~\cite{myers13}, was very successful and resulted in important discoveries in physics. In spring 2013, the LHC was shut down for about 2~years, in order to allow consolidation of the superconducting splices in the magnet interconnects, following the incident of 2008. In parallel, numerous other machine systems have been consolidated or upgraded. A common goal of the upgrades is to improve the machine so that it can safely operate closer to its design energy and thus extend the physics discovery potential. For the restart of the LHC in 2015, several challenges can be anticipated, and it is important to carefully define its operational parameters at the start-up in order to maximize the chances of a smooth and successful second running period. 

In this paper, we discuss first the general strategy for 2015, which leads up to a proposed choice of starting configuration. Our focus is on collimator settings and reach in \bs, since most other parameters are covered by other papers in these proceedings~\cite{solfaroli14_cham,meschi14_cham,wenninger14_cham,iadarola14_cham,yannis14_cham,kain14_cham,butterworth14_cham,massimo14_cham}. We discuss also how the performance can be increased later in the run, when the operational behavior of the machine is better known. 

\section{strategy for 2015}
When the LHC restarts in 2015, it will operate at a higher energy and shorter bunch spacing than in 2012 (6.5~TeV and 25~ns compared to 4~TeV and 50~ns)~\cite{solfaroli14_cham,meschi14_cham}. These changes imply new major operational and beam physics challenges. Furthermore, the higher beam energy and potentially larger total beam intensities make the LHC beams more dangerous. Fewer protons are needed to cause quenches or damage of sensitive machine components. At the same time, the risk of a known serious failure mode, the asynchronous beam dump, increases at higher energy~\cite{magnin14_evian}, and a higher rate of UFOs is expected~\cite{auchmann14_cham}. It is also uncertain how the operational issues encountered in 2012, such as instabilities and beam lifetime drops, will be manifested at 6.5~TeV.

Because of the many uncertainties, the operational behavior of the machine in 2015 is not as well known as in the end of Run~I, which means that the beam commissioning risks to be less smooth as in 2012. Therefore, we envisage in the operational strategy for 2015 a careful start of the LHC in a relaxed configuration, which allows larger operational margins. The focus is put on feasibility, stability, and ease of commissioning, and the main priority is not peak performance but rather to establish a running machine at 6.5~TeV and 25~ns. Where possible, it should be avoided to introduce too many new features at once. On the other hand, the starting parameters should also not be overly pessimistic. Therefore, the operational achievements in Run~I are used, where possible, to deduce what is likely to work. 

The main focus in this paper is to define the machine parameters for the start-up, but we discuss also, at the end of the paper, what changes can be made later in the year. Once sufficient beam experience is gathered through machine development sessions~\cite{LSWGday_2014} or routine operation, the luminosity performance could be pushed. The ultimate reach in luminosity is hard to predict but we give an overview of the different parameters that can be adjusted.

Even though the final goal is to operate at 25~ns, a short initial run will take place at 50~ns. In order to save commissioning time, this run will use the same machine configuration as the 25~ns run. Therefore, we do not discuss in further detail the 50~ns run. 

\begin{figure}[tb]
  \centering
\includegraphics[width=8.5cm]{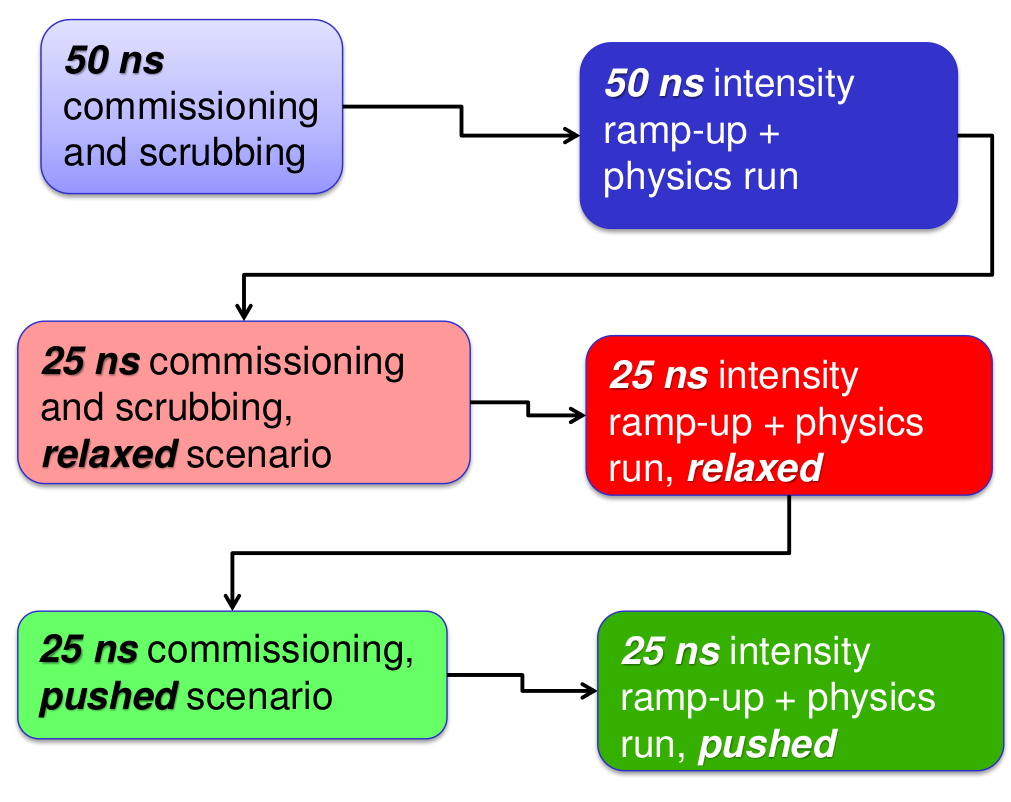}
  \caption{Schematic illustration of the different proton run periods envisaged for the LHC in 2015.}
\label{fig:lhc2015_outline}
\end{figure}

These different stages of the 2015 proton physics period are schematically summarized in Fig.~\ref{fig:lhc2015_outline}. Each physics run has to be preceded by a scrubbing period to mitigate the effects of electron cloud~\cite{iadarola14_cham} and possibly by additional commissioning. 

Further details of the 2015 run can be found in Ref.~\cite{wenninger14_cham}.

\section{Beam characteristics}

Although the design proton beam energy of the LHC is 7~TeV, the baseline energy for 2015 is 6.5~TeV. The reason is that, in order to reach 7~TeV, it is estimated that an unfeasibly large number of training quenches is needed~\cite{ezio12_chamonix}, although this estimate might be adjusted in the future, when more results of powering tests become available~\cite{solfaroli14_cham}. 

There has been a strong request from the experiments to operate with the design bunch spacing of 25~ns, since it provides potentially higher luminosity and lower pileup~\cite{meschi14_cham}. The 25~ns scheme is, however, coupled to several potential complications, for example stronger electron cloud~\cite{iadarola14_cham} and the need of a larger crossing angle to compensate for the stronger long-range beam-beam effect~\cite{pieloni14_evian}.

The characteristics of the LHC bunches in physics operation are strongly dependent on the beam provided by the injectors. Presently, the injectors can provide two different types of beams: BCMS (batch compression and merging and splittings) and nominal~\cite{yannis14_cham}. In both schemes, the achievable bunch intensity is, under optimistic assumptions, up to about $1.3\times10^{11}$, which is slightly higher than the nominal $1.15\times10^{11}$. The BCMS beams have significantly smaller emittances (at LHC injection down to 1.3~\um normalized emittance compared to 2.4~\um for nominal) but fewer bunches (2544 or 2592 colliding in IR1 and IR5, depending on the number of trains, compared to 2736 for nominal).

Once injected in the LHC, intensity loss and emittance growth are very likely to occur. Using typical numbers from Run~I, an intensity loss of about 5\% could be expected, which leaves a bunch intensity up to about $1.2\times10^{11}$ in collision. The emittance is affected by several physical processes. If only the unavoidable effect from intrabeam scattering (IBS) is accounted for, growths of 20\% or 5\% have been calculated~\cite{kuhn14_evian} for BMCS and nominal respectively. However, if the scrubbing runs are not fully successful in mitigating the electron clouds, a much larger emittance growth is likely to occur~\cite{iadarola14_cham}.

Although with a potentially much higher peak luminosity, it is not obvious that BCMS is the better choice, since the very small emittance could have a detrimental effect on the single-beam stability~\cite{mounet14_evian}. In addition, the small emittances are more challenging for machine protection~\cite{kain14_cham}. Therefore, the choice between the two beams is still at the time of writing (September 2014) an open question.

In the longitudinal plane, a bunch length of 1.25~ns can be expected at injection and 1.2~ns in collision, for the RF voltages of 6~MV (injection) and 12~MV (collision)~\cite{muller14_evian,butterworth14_cham}. Shorter bunches of nominal length (about 1~ns) could be within reach from the machine side and could be put into operation. Possibly, the increased pileup  density can be handled by the experimental detectors~\cite{meschi14_cham}. A shorter bunch length would be beneficial for the luminosity since the geometric reduction factor is increased.

\section{LHC cycle and optics}
Several significant changes to the LHC operational cycle are under study. Examples of such changes are luminosity leveling by dynamically changing \bs during stable beams (in order to reduce the pileup), putting the beams into collision already before the squeeze starts (in order to stabilize the beams in the squeeze using the tune spread introduced by the collisions)~\cite{metral_LMC_14} or combining the ramp and the squeeze (to make the cycle shorter)~\cite{ipac12_redaelli_ramp_squeeze}. 
With the philosophy that it should be avoided, where possible, to introduce untested features at the 2015 start-up, these operational improvements are a priori not a part of the start-up baseline, but could instead be introduced at a later stage in the run when more experience has been gained. A detailed account of the nominal cycle is given in Ref.~\cite{solfaroli14_cham2}.

Two different optics schemes have been under consideration: the nominal optics~\cite{lhcdesignV1}, used in Run~I, and the achromatic telescopic squeeze (ATS)~\cite{fartoukh13_prstab}. ATS is a promising option that could provide several advantages, but it has also some outstanding points that need further study~\cite{LMC_bs_2014}. Therefore, it has been decided to start with nominal optics, while keeping the possibility to switch to ATS at a later point. Further details are given in Ref.~\cite{massimo14_cham}.

\begin{table*} \centering
  \caption{Settings of different collimator families for different scenarios for 6.5 TeV operation after LS1, where either the 2012 settings are kept in~mm, in \sig or more open (relaxed). All settings are given in units of the local transverse beam size \sig, which is calculated using the $\beta$-function at each collimator and the nominal emittance of 3.5~\um. \vspace{2mm}}
  \label{tab:coll_settings}
\begin{tabular}{|l|r|r|r|} \hline
  Collimators  & Relaxed settings (\sig)  & mm settings kept (\sig) & $\sigma$ settings kept (\sig) \\  \hline \hline
  TCP7   & 6.7 & 5.5 & 5.5  \\ \hline
  TCS7  & 9.9 & 8.0 & 7.5  \\ \hline
  TCLA7  & 12.5 & 10.6 & 9.5  \\ \hline
  TCS6  & 10.7 & 9.1 & 8.3 \\ \hline
  TCDQ6  & 11.2 & 9.6 & 8.8 \\ \hline 
  TCT  & 13.2 & 11.5 & 10.7 \\ \hline 
  protected aperture (\sig) & 14.8 & 13.4 & 12.3 \\ \hline
  \end{tabular}
\end{table*}

The injection optics for 2015 will thus stay the same as in 2012. At top energy, a new final point of the \bs squeeze has to be decided upon, together with a new crossing angle. This is discussed in detail in the following sections for IR1 and IR5, where \bs is limited by the available aperture. In IR2 and IR8, \bs is instead adjusted to the luminosity that the detectors can handle, and the aperture is less critical. IR2 will therefore stay at the injection value of \bs=10~m with an external half crossing angle of 120~\urad, while IR8 will use the same configuration of \bs=3~m as in 2012 and an external half crossing angle of -250~\urad. It should be noted though that the crossing angles in IR2 and IR8 are under review by the ABP/HSC section to ensure that beam-beam effects are in the shadow of IR1 and IR5. In all IRs except IR8, a parallel separation of 2~mm will be used at injection, as in 2012. In IR8, the parallel separation has to be increased to 3.5~mm  with the addition of a parallel angle of 40~\urad~\cite{fartoukh_LMC_13}. In collision, the 2012 value of the parallel separation is rescaled by the energy and rounded to obtain 0.55~mm at all IRs.

\section{Collimator settings}
The LHC collimation system~\cite{lhcdesignV1,assmann05chamonix,assmann06,bruce14_PRSTAB_sixtr} influences directly the peak luminosity performance in several ways. Firstly, the cleaning inefficiency (the local losses in a cold element normalized by the total losses on collimators), together with the beam lifetime and the quench limit, define the maximum acceptable intensity. Secondly, when pushing the \bs to smaller values, the $\beta$-function in the inner triplets increases, meaning that the normalized aperture margin between the central orbit and the mechanical aperture decreases. If this margin becomes too small, the aperture can no longer be fully protected by the collimation system. At what aperture this occurs depends on the collimator settings. The loss in aperture is further enhanced by the fact that a larger crossing angle is needed at smaller \bs in order to keep the same normalized  beam-beam separation. The collimators are also the main contribution to the LHC impedance, which is crucial for beam stability.

Different collimator settings have been under consideration for the 2015 start-up and the three main scenarios are shown in Table~\ref{tab:coll_settings}. In terms of cleaning, the relaxed settings are close to the limit of preventing a beam dump at a beam lifetime of 12~minutes and full nominal intensity, even though significant uncertainties exist~\cite{verweiCollReview2013}. Although a detailed verification with final optics is pending at the time of writing, it is expected that the other two types of settings have better cleaning efficiency that should suffice, unless the beam lifetime drops significantly below the 12~minutes specification, or the quench limit would be much worse than expected. Therefore, we do not expect the cleaning inefficiency to be a limiting factor for the total beam intensity.

In order to be on the safe side for the cleaning, but without going to the tighter gaps with the 2~\sig retraction that are more challenging in terms of impedance, Run~II will start with the 2012 settings kept in~mm (middle column in Table~\ref{tab:coll_settings}). They also have a well-proven long-term stability in terms of preserving the hierarchy under unavoidable drifts of optics and orbit. 

The impedance and single-beam stability for the different collimator settings are discussed in Ref.~\cite{mounet14_evian}. It is shown that for the nominal, large-emittance beam, all proposed collimator settings should provide sufficient stability with both octupole polarities, while stability could be an issue with the BCMS beams. The two-beam effects and octupole polarities are discussed in detail in Ref.~\cite{pieloni14_evian}. 

For machine protection, the settings in Table~\ref{tab:coll_settings} fulfill the same demands as used during Run~I~\cite{bruce11evian,ipac11_bruce_betaStar} in terms of the IR6 dump protection shadowing the tertiary collimators (TCTs) and the TCTs shadowing the triplets. The margins between different collimator families are calculated based on what was achieved in Run~I. If the stability of the optics or orbit correction for post-LS1 would be worse, larger margins are needed. Furthermore, the TCT damage limit in number of protons is lower and the baseline 25~ns filling scheme means that there is a risk to have double the number of bunches within the critical time window during asynchronous dumps when bunches pass the dump kickers and receive intermediate kicks. Therefore, it could be wise to introduce more margins at the start-up, before the machine stability is well known, in order to be sure that the TCTs and aperture are protected. This is especially true at more relaxed values of \bs, where the orbit in~mm scales to a larger variation in units of \sig so that larger margins in \sig are needed. 

For the calculation of \bs we first investigate the limiting configuration with a protected aperture of 13.4~\sig from Table~\ref{tab:coll_settings} and then evaluate more relaxed scenarios.

\section{Aperture and crossing angle}
Given the aperture that is protected by the collimation system, the achievable \bs can be calculated, if also the required aperture as a function of \bs is known. This function depends both on which tolerances are included in the aperture calculation and on the required crossing angle. 

The aperture was measured during Run~I on several occasions~\cite{pons10, redaelli11_apertureIR15, ipac11_assmann_aperture, ipac12_redaelli_aperture, redaelli12_IR2aperture, hermes13_IR8aperture, hermes13_IR2aperture}, using the circulating beam, and it was found that the results were compatible with a very well aligned machine with very small errors~\cite{bruce14_n1_ap_meas}. During the shutdown, all magnets have been realigned, so the alignment should  a priori not be worse than at the start of Run~I.

For the aperture calculation, we therefore assume that the aperture has not become worse during LS1 and, at this stage, we do not include additional safety margins on orbit or optics. However, we base our calculations on the most pessimistic measurement from 2012. We scale this measured value by \bs and add the change in orbit due to a different crossing angle, in order to estimate the crossing plane aperture at any other configuration. This straightforward, analytic method has proven to give results very close to the MAD-X aperture model~\cite{bruce14_evian}.

To verify the calculations, it is very important that the aperture is measured with beam very early on during the commissioning, after the reference orbit has been established and the optics corrected. If it turns out that the assumptions were too optimistic, the time loss when stepping back to a larger \bs, if needed, should be very small. 

The criteria for choosing an appropriate crossing angle for 2015 are discussed in Ref.~\cite{pieloni14_evian}. It needs to be sufficiently large to minimize the detrimental effects of the long-range beam-beam interactions, but when the angle is increased, the available aperture margin goes down. In order to calculate the needed crossing angle as a function of \bs, Ref.~\cite{pieloni14_evian} suggests to use a normalized beam-beam separation of 11~\sig for the start-up, based on simulations of dynamic aperture and operational experience from Run~I. The larger-than-nominal separation is motivated by the possibility to have also larger intensities, e.g. $1.3\times10^{11}$ protons per bunch. 

In the calculation, a normalized emittance of 3.75~\um is used. If the real beam would have a smaller emittance, the calculated crossing angles in \urad still provide sufficient beam-beam separation.

\section{\bs at start-up}

\begin{figure*}[tb]
  \centering
\includegraphics[width=12cm]{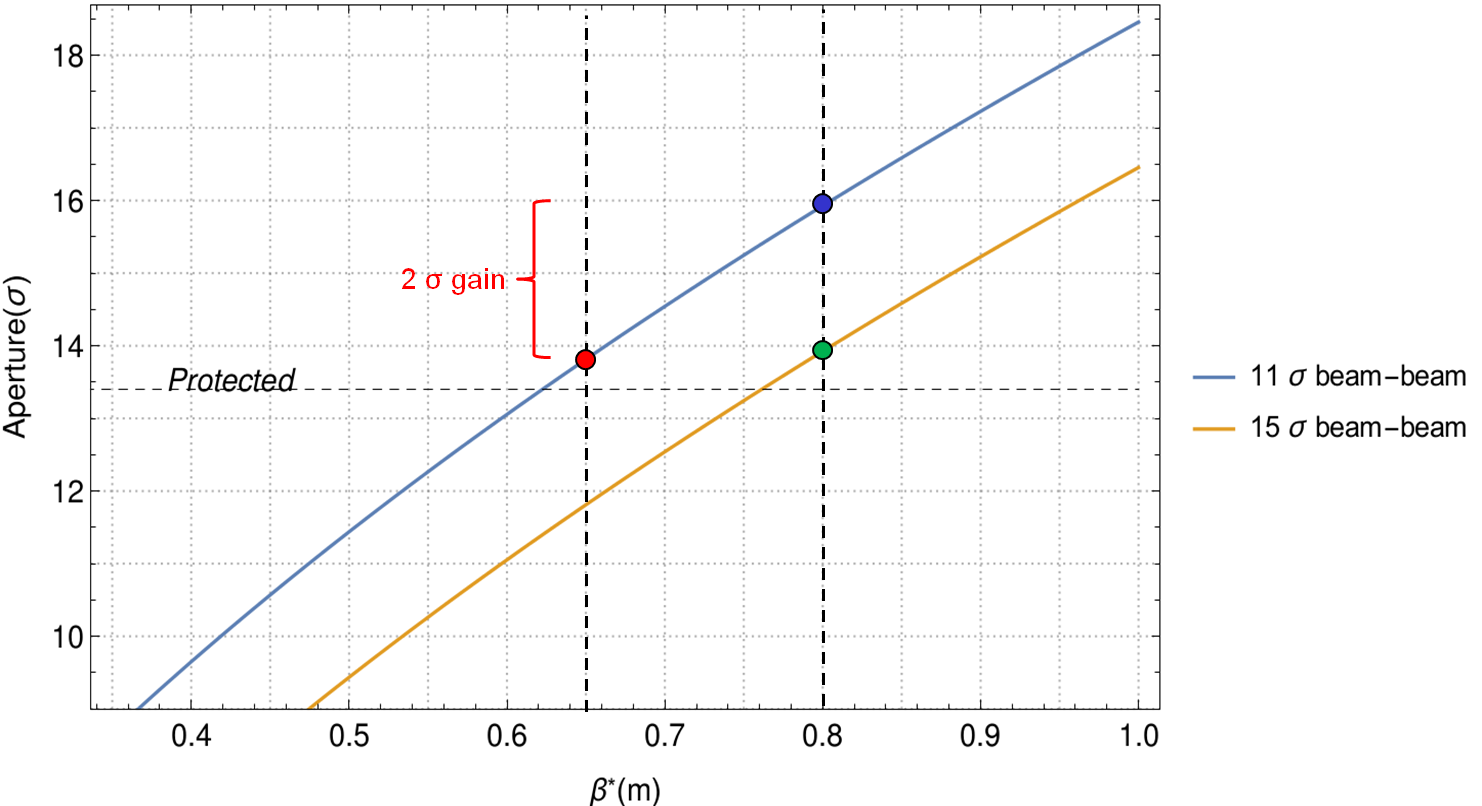}
  \caption{Estimated aperture as a function of \bs, assuming 11~\sig (blue line) or 15~\sig (yellow line) beam-beam separation. As the beam-beam separation is kept constant, the crossing angle in $\mu$rad varies along each curve. The dashed horizontal line shows the minimum protected aperture if the 2012 collimator settings in mm are kept (see Table~\ref{tab:coll_settings}. The red dot shows the limiting \bs=65~cm with this protected aperture, rounded to a matched 5~cm point. It corresponds to a half crossing angle of 160~\urad. The blue dot shows the baseline operating configuration at \bs=80~cm, and a half crossing angle of 145\urad, where the beam-beam separation of 11~\sig is kept. The green dot shows a possible operating configuration at \bs=80~cm and a half crossing angle of 195\urad.}
\label{fig:ap_vs_bs}
\end{figure*}

\begin{figure}[tb]
  \centering
\includegraphics[width=8cm]{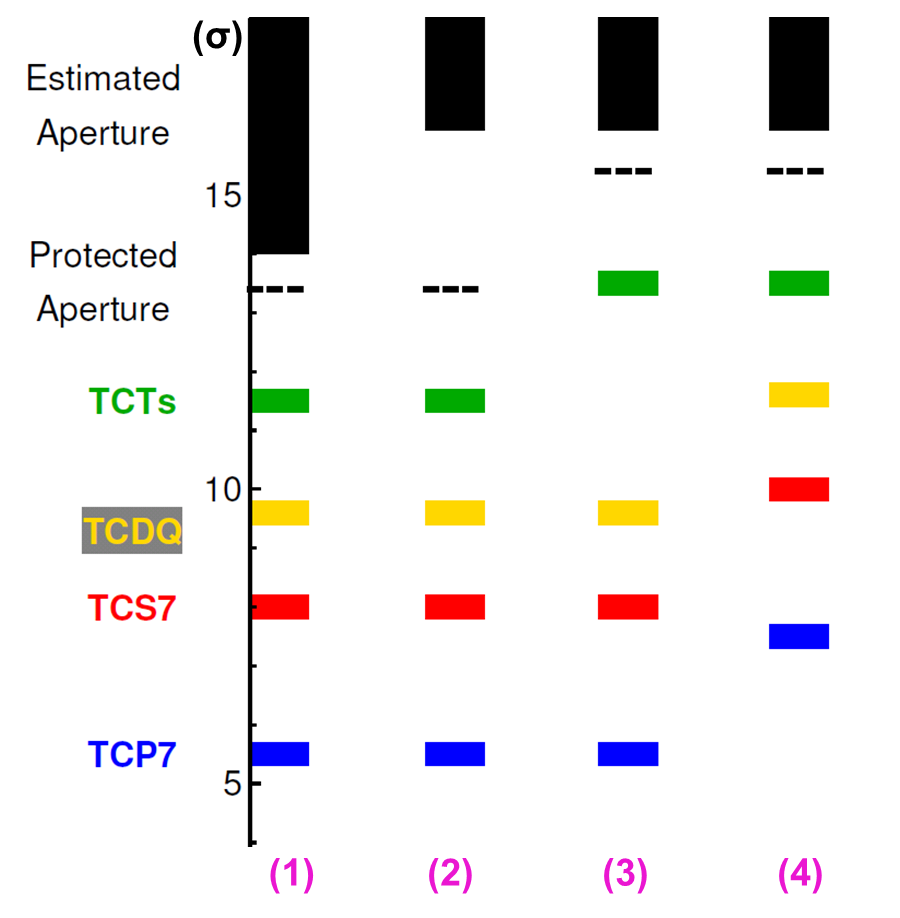}
  \caption{Schematic illustration of the margins in the collimation hierarchy, starting from a limiting configuration (1) at \bs=65~cm. In (2) the available aperture is larger but no collimator moved. In (3) the TCT is moved out, in order to be better protection against asynchronous dumps,  and in (4) the whole collimation hierarchy is moved out to reduce the impedance. }
\label{fig:margin_use}
\end{figure}

The required aperture as a function of \bs is shown in Fig.~\ref{fig:ap_vs_bs}, assuming a constant beam-beam separation of 11~\sig (blue line). Under the assumption that the protected aperture is 13.4~\sig, and that we operate at points rounded to a 5~cm spacing, the limiting \bs that could be achieved is 65~cm (illustrated by the red dot in Fig.~\ref{fig:ap_vs_bs}). This configuration, corresponding to a 160~\urad half crossing angle, has been discussed in detail in Ref.~\cite{bruce14_evian}. It should be noted that the rounding up to 65~cm introduces a small aperture margin---the aperture prediction has anyway an error margin not smaller than the measurement precision of 0.5~\sig.

It should be noted that several of the underlying assumptions on protection and stability contain uncertainties. For example, it cannot be guaranteed a priori that the orbit stability and optics correction will be as good as in 2012. Furthermore, the scaling to higher energy of instabilities and lifetime drops, presumably connected to the collimator impedance, is not known with a high accuracy. Therefore, in view of the approach of a relaxed start-up, it is wise not to start at the limiting configuration, but instead allow some additional margins. 

Based on these considerations, it has been decided to start the 2015 LHC run at \bs=80~cm~\cite{LMC_bs_2014}. If the beam-beam separation is kept constant at 11~\sig, the baseline operating configuration is therefore the blue dot in Fig.~\ref{fig:ap_vs_bs}, where a half crossing angle of 145\urad is found. It can be seen in the figure that the step to \bs=80~cm frees about 2~\sig of aperture margin, which could be used in different ways depending on where it turns out to be needed. 

If no collimators are moved, the additional margin just increases the aperture budget and makes it more certain that the real measured aperture will be compatible with the protected one. This is illustrated schematically in steps (1) and (2) in Fig.~\ref{fig:margin_use}. 

In order to compensate for the uncertainty in orbit stability and optics correction, as well as the higher risk of asynchronous dumps at 6.5~TeV, the margin can be used to move out the TCTs so that they are better protected, as shown in step (3). 

Step (4) in Fig.~\ref{fig:margin_use} illustrates yet another possibility, where all collimators are moved out in order to reduce the total machine impedance. This option could be envisaged if the beam stability turns out to be limited by impedance effects. A similar option, where all collimators but the primary (TCP) are moved out, could also be envisaged. This option would allow a learning curve for loss spikes with small TCP gaps. 

In case the long-range beam-beam tune shift would turn out to be limiting, the additional aperture margin could also be used to increase crossing angle. This is illustrated by the green dot in Fig.~\ref{fig:ap_vs_bs}. As can be seen, if all additional margin would be dedicated to the beam-beam separation, it could be increased to about 15~\sig at \bs=80~cm. This configuration corresponds to a half crossing angle of 195~\urad. 

It is not yet decided which of the different options for using the additional margin that will be used. One could also use a split between several of them. The partition of the margins could even be changed during the 2015 commissioning, when it is clearer where it is mostly needed, although some changes would require additional commissioning time. We list here some examples of realistic suggestions for the start-up:
\begin{itemize}
 \item All margin on machine protection: This option compensates for uncertainties on failure probabilities and, with the 11~\sig beam-beam separation and tight collimators, it allows us to learn early on about potential limitations on beam stability.
 \item 1~\sig on machine protection and 1~\sig on beam-beam separation: This option allows a more relaxed squeeze with lower probability of instabilities, while maintaining a higher level of protection. It should be noted that 1~\sig of aperture translates approximately into 2~\sig of beam-beam separation, meaning a total separation of 13~\sig and a half crossing angle of 170~\urad.
\end{itemize}

\section{Ways to push performance}
Once the LHC has been successfully put into operation and a first period of stable beams has been established, it is reasonable to assume that the performance limitations will be better known. Then, the performance could be increased based on the operational experience and possible MDs. Several machine parameters could be changed to gain in luminosity performance:
\begin{itemize}
 \item 	Bunch intensity: As the peak luminosity depends on the square of the bunch intensity, increasing it is a very efficient (and well-known) way to boost the performance. The intensity is mainly limited by the performance of the LHC injectors~\cite{yannis14_cham} and by the beam stability in the LHC~\cite{pieloni14_evian,mounet14_evian}.
 \item Smaller emittance: This is also a well-known and straightforward way to increase the luminosity. It is also limited by the injectors and beam stability, but also by machine protection considerations~\cite{kain14_cham}. It introduces also an additional gain by allowing a smaller crossing angle in \urad and therefore a larger aperture margin.
 \item Collimator settings: If the margins in the hierarchy are reduced, e.g. by establishing the 2~\sig retraction settings in Table~\ref{tab:coll_settings}, a smaller aperture can be protected, and thus a smaller \bs tolerated. However, with tighter settings, the impedance increases. Whether this is tolerable has to be evaluated with beam. Based on further operational experience, the margins between the dump protection and the TCTs, as well as the margins between TCTs and triplets, might be decreased if the new integrated BPM buttons can be used to reduce orbit drifts from the center of the collimators. The less temperature-sensitive BPM electronics could also be used to determine whether some of the large orbit drifts between TCTs and triplets, observed in Run~I, are real or an artifact of the measurements. In the future, we still hope to achieve nominal collimator settings in IR7 with a 1~\sig retraction between the TCP and the secondary collimators (TCS). However, because of the impedance constraints, this is unlikely to be usable during Run~II. Installing new TCSs made of materials with lower impedance could help. Furthermore, integrated BPMs in the TCS would help to ensure that the hierarchy is maintained in spite of the smaller margin.
 \item Crossing angle: reducing the crossing angle at a given \bs implies a gain in the required aperture. However, if the beam-beam separation is decreased, the long-range effect becomes more critical, in particular during the squeeze~\cite{pieloni14_evian}, which limits the smallest achievable crossing angle.
 \item Aperture: unless additional margins are introduced at the start-up, the gain should be rather small. The aperture in Run~I was found in measurements to be very close to the ideal one, and the same assumptions are used for Run~II.
 \item Bunch length: with a shorter bunch length, the geometric reduction factor is closer to one and the luminosity loss smaller. A shorter bunch length is likely to be within reach from the machine side~\cite{muller14_evian,butterworth14_cham} and could possibly be put in place.
\end{itemize}

We cannot a priori determine the exact limit of actual \bs-values that could be reached later in the run, as many underlying parameters must be examined with beam. Instead, we give a few examples of possible configurations with pushed performance:
\begin{itemize}
 \item \textbf{ $\boldsymbol{\beta}^*$=65~cm:} From Fig.~\ref{fig:ap_vs_bs} it is clear that \bs=65~cm could be within reach even with rather conservative assumptions. 
 \item  \textbf{ $\boldsymbol{\beta}^*$=55~cm:} If beam studies show that the impedance is acceptable for reduced collimator settings with a 2~\sig retraction in IR7 (see Table~\ref{tab:coll_settings}) \bs=55~cm could be within reach if the aperture is at the limit of what can be tolerated. Alternatively, the main gain could come from the crossing angle. Keeping the mm~kept settings, \bs=55~cm and a crossing angle of 130~$\mu$rad fits almost exactly within the protected aperture. This configuration corresponds to a beam-beam separation of 8.3~\sig for an emittance of 3.75~$\mu$m. If the emittance can be reduced to 2.5~$\mu$m, the beam-beam separation with this crossing angle is about 10~\sig. This configuration is possibly compatible with 6~\sig dynamic aperture~\cite{pieloni14_evian} but the limit would have to be deduced from beam studies.
 \item \textbf{ $\boldsymbol{\beta}^*$=40~cm:}  This configuration could be within reach under optimistic assumptions~\cite{bruce14_evian}. For this ultimate scenario for Run~II we assume the 2~\sig retraction collimator settings, with the addition of using the BPM button collimators to their full potential. Furthermore, we assume a beam-beam separation of 10~\sig at an emittance of 2.5~$\mu$m. These assumptions are considered challenging but possible, although it is not given that this configuration can be used. It could also require significant beam experience and additional commissioning time. Based on the possibilities of reaching \bs=40~cm, the optics will be commissioned down to this value already at the start-up, in order to have maximum flexibility. As an alternative to round optics, the configuration with \bs=40/50~cm in the two planes might be easier to reach in terms of aperture and gives comparable luminosity. 
\end{itemize}

\section{Conclusions and outlook}
The LHC will be re-started in 2015 after about two years of shutdown. Many hardware changes and upgrades have taken place and the machine will operate at a higher energy of 6.5~TeV energy and a shorter bunch spacing of 25~ns. Therefore, the machine behavior is less well known than at the end of Run~I and the strategy for 2015 is to start carefully with the main aim to get the machine running safely and stably. 

Based on these considerations, we have presented the LHC baseline parameters for the 2015 start-up, which we summarize for convenience in Table~\ref{tab:par}. Most notably, the LHC will start proton physics at \bs=80~cm, a 145~\urad half crossing angle, and 2012 the collimator settings kept in mm. It is at the time of writing not decided whether the nominal or BCMS beams from the injectors will be used. These parameters contain some margins which could be used for increased machine protection, or, in case of need, for a relaxed beam-beam separation or impedance.


Later in the run, a push in \bs and performance can be envisaged, when the operational limits are well established based on beam experience. This pushed \bs-value could be as low as 40~cm under optimistic assumptions. 

\begin{table*} \centering
  \caption{Summary of the main LHC beam and machine parameters for 2015. It should be noted that the emittance values in collision are optimistic and assume emittance growth only from IBS with values from Ref.~\cite{kuhn14_evian}. If the scrubbing is not fully successful, larger emittances should be expected. Furthermore, the intensity in collision assumes a 95\% transmission of the injected intensity. It should also be noted that the 2012 mm kept collimator settings in collision might still be modified to achieve a larger margin for machine protection between the TCDQ and the TCTs. \vspace{2mm}}
  \label{tab:par}
\begin{tabular}{|l|c|r|r|} \hline
  Parameter  & Unit & Value at injection & Value at collision\\  \hline \hline
  Beam energy & TeV & 0.45 & 6.5 \\ \hline
  \bs at IR1/IR2/IR5/IR8 & m & 11 / 10 / 11 / 10 & 0.8 / 10 / 0.8 / 3 \\ \hline
  half crossing angle at IR1/IR2/IR5/IR8 & \urad & -170 / 170 / 170 / 170 & -145 / 120 / 145 / -250 \\ \hline
  Tunes (H/V) & -- & 64.28/59.31 & 64.31/59.32 \\ \hline
  Parallel separation at IR1/IR2/IR5/IR8 & mm & 2 / 2 / 2 / 3.5 & 0.55 / 0.55 / 0.55 / 0.55 \\ \hline
  Normalized emittance (BCMS/nominal) & \um & $\geq 1.3$ / $\geq 2.4$ & $\geq 1.7$ / $\geq 2.7$ \\ \hline
    Total number of bunches (BCMS/Nominal) &-- & \multicolumn{2}{|c|}{$\leq 2604$ / 2748 } \\ \hline
  Number of bunches colliding at IR1/5 (BCMS/Nominal) & -- & \multicolumn{2}{|c|}{ $\leq 2592$ / 2736 }\\ \hline
  Bunch intensity  & p & $\leq 1.3\times10^{11}$ &  $\leq 1.2\times10^{11}$ \\ \hline
  Bunch length (4\sig) & ns & 1.0--1.2 &  1.0--1.25 \\ \hline
  Collimator settings & -- & 2012 mm kept &  2012 mm kept \\ \hline \hline
  \end{tabular}
\end{table*}

\section{ACKNOWLEDGMENTS}
The authors would like to thank numerous colleagues for input and discussions: H. Bartosik, X. Buffat, E.~Chapochnikova, P. Collier, R. de Maria, G. Iadarola, V. Kain, E. Meschi, N. Mounet, Y. Papaphilippou, G. Papotti, G. Rumolo, B. Salvachua, B. Salvant, M. Solfaroli, R.~Tomas, G.~Valentino, D. Valuch, and M. Zerlauth



\end{document}